# Digitally Mutating NV-FPGAs into Physically Clone-Resistant Units

Ayoub Mars and Wael Adi

*Abstract*— The concept of Secret Unknown Ciphers SUCs was introduced a decade ago as a new visionary concept without devising practical real-world examples. The major contribution of this work is to show the feasibility of "self-mutating" unknown cipher-modules for physical security applications in a non-volatile FPGA environment. The mutated devices may then serve as clone-resistant physical units. The mutated unpredictable physical digital modules represent consistent and low-cost physical identity alternatives to the traditional analog Physically Unclonable Functions (PUFs). PUFs were introduced two decades ago as unclonable analog physical identities which are relatively complex and suffer from operational inconsistencies. We present a novel and practical SUC-creation technique based on pre-compiled cipher-layout-templates in FPGAs. A devised bitstream-manipulator serves as "mutation generator" to randomly-manipulate the bitstream without violating the FPGA design rules. Two large cipher classes (class-size larger than $2^{1000}$) are proposed with optimally designed structure for a non-volatile FPGA fabric structure. The cipher-mutation process is just a simple random unknown-cipher-selection by consulting the FPGA's internal True Random Number Generator (TRNG). The security levels and qualities of the proposed ciphers are evaluated. The attained security levels are scalable and even adaptable to the post-quantum cryptography. The hardware and software complexities of the created *SUCs* are experimentally prototyped in a real field FPGA technology to show very promising results.

## I. Introduction

THE majority of contemporary systems require physically hard to clone (non-replaceable) modules as security anchors. Physical(ly) Unclonable Functions (PUFs) were proposed to serve as such anchors [1][2][3]. PUFs were deployed in many applications such as secure memoryless key storage [4] and as unclonable entity identifiers [5][6]. However, PUFs drawbacks and their high complexity limited so far, their practical usage especially in automotive and consumer mass-products.

Emerging VLSI self-reconfiguration capabilities inspired the authors a decade ago to introduce a concept for self-creating "electronically-mutating" irreversibly unknown hard-wired ciphers in VLSI devices in a post-fabrication process [7]. The reader may wonder whether an unknown cipher in a device makes sense at all. Or ask fundamentally about the use of ciphers which nobody knows? Many use-cases for such ciphers were shown in making devices physically unclonable or clone-resistant for a large class of IoTs applications [8] [9] [10].

This work demonstrates for the first time one not ultimate however, possible practical approach towards creating such ciphers in real-field non-volatile FPGA devices. The expected possible realization approaches for unknown ciphers seem to be unlimited. Admittedly, FPGA fabrics allowing even the proposed approach do not contemporarily exist however are expected to emerge soon. This work was presented first in [11].

As the concept of unknown ciphers is not well known in the common literature, and to make the paper self-contained, the key ideas of the visionary unknown cipher concept are summarized in section II.

**Contributions:** The <u>main</u> <u>contributions</u> of this work can be summarized as follows: *First*, showing that creating unknown ciphers in VLSI technology is a feasible task. *Second*, a novel and efficient concept for embedding a practical self-reconfiguring-manipulator in SoC FPGAs is proposed. It is based on creating cipher-templates in the fabric and NV-memory and internal bitstream manipulators without violating the FPGA design rules. *Third*, two new large classes (class size $\approx 2^{1234}$ and $2^{1350}$) of random ciphers adapted to best-fit into the non-volatile SoC FPGAs fabric structure are presented. The created cipher in a device is considered as unknown as an internal true random generator is ensuring to select just one unknown choice out of $2^{1234}/2^{1350}$ ciphers. *Fourth*, the resulting unknown ciphers are shown to be secure against most classical attacks and being scalable even for post-quantum cryptography with adequate complexity. The implementation complexity is evaluated by sample prototyping in Microsemi SmartFusion®2 SoC FPGA technology, showing the feasibility and efficiency of the concept and the required FPGA bitstream controller changes to allow such self-creation process.

## II. The New Paradigm of Mutating Unknown Ciphers

The term "Unknown-Cipher" seems for the first moment as a contradiction, as at least the one who designed the cipher should know it? The authors postulated that the emerging VLSI technology would allow practical self-creation of unknown entities and introduced a first visionary bio-inspired mutation process within VLSI devices that was called a "digital mutation" in [7]. Such intended mutations (hypermutations for immunity), should allow creating permanent *unknown cryptographic entities such as secrets ciphers* or *hash* functions etc. within VLSI devices [12][13] in a post-fabrication process. To attain such usable unknown functions, a smart infrastructure with self-reconfiguration capability in non-volatile (NV) FPGA

Ayoub Mars and Wael Adi are with the Institute of Computer and Network Engineering, Technical University of Braunschweig, 38106 Braunschweig Germany (e-mails: a.mars@tu-bs.de, w.adi@tu-bs.de ) environment is then required. Such technologies do not yet really exist. However, are expected to emerge in a near future VLSI devices. The following key-concepts are presented as backgrounds for the secret unknown cryptography.

*A. Unknown-Ciphers Paradigm and Kerckhoffs's Principle*

A traditional secret is something known to somebody as a privilege allowing that person to have exclusive access rights. Such secret can be willingly duplicated and forwarded to another entity to share that privileged access. If a secret is not known like a Physical Unclonable Function (PUF), then it should be physically forwarded to allow privilege sharing. In that case, the privileged-owner would lose his or her access rights. PUFs are born natural properties which are somehow equivalent to *weak unknown hash functions*. To our knowledge, there are no *natural physically born unknown ciphers* (permutations) such as PUFs. Our visionary concept in unknown cryptography is to create unknown ciphers. The concept of unknown ciphers is not to be confused with "obscured ciphers". Therefore, the proposed unknown cipher or possibly unknown cryptographic functions should not lead to "security by obscurity". Kerckhoffs's principle assumes that any used ciphers are impossible to be kept secret and the only secrets are the keys shared by both communicating parties.

We postulate that *the only perfect secret or cryptographic-function is the one which nobody knows or capable to predict.* Therefore, attacking such secrets is equivalent to an exhaustive search over the whole possible keys and/or functions space. We expect that VLSI technology would reach soon the capability to create practical entities incorporating "unpredictable" unknown cryptographic functions. We start with the attempt to set concepts for creating secret unknown ciphers SUCs.

*A Secret Unknown Cipher SUC:* If a cipher is designed by a cryptographer and is then kept secret in production, then this leads to the typical case of "Security by Obscurity". Such cases failed so far practically in all known applications as the security concepts violate Kerckhoffs's principle. However, if the cipher creator/generator itself cannot predict the generated cipher, then the cipher is deemed as not known.

*Bounds on Unknown Ciphers*: The unpredictability of a designed cipher is upper bounded. A secret unknown cipher (SUC) of *n-to-n bits* is seen to be perfectly created, if it is a non-predictable (unknown) choice out of all possible $2^n!$ *n-bit* permutations. To get an idea about the huge space of that cipher-choice for a small *n* as *n=10*. The number of all possible invertible 10-to-10 bits mappings including trivial cases is $2^n! = 1024! \approx 2^{8192}$ *choices.*

<u>*Interesting VLSI-FPFA-specific practical examples:*</u>

1. *Constructing four 4-LUTs as (4-to-4 bits mapping) allows a total of $2^{64}$ possible mappings. Only $2^4! \approx 2^{44}$ of them are invertible permutation functions or ciphers including all trivial cases. About $2^{20}$ of them are known as useful optimal 4-bit S-Boxes for crypto mappings!*
2. *A 6-to-6-bit mapping allows $2^6! \approx 2^{296}$ possible invertible mappings/ciphers which may be accommodated within a very small FPGA cell structure.*
3. *A traditional cipher of 64-to-64 bits has a cardinality space of about $2^{64}! \approx 2^{2^{70}}$ possible ciphers.*

*B. Are Secret Unknown Ciphers (SUCs) Creatable?*

SUC creation seems to be a very challenging task however, not impossible. One objective of this paper is to show the feasibility of their practical realization in view of technology progress.

Assume that a NV FPGA would allow internal self-reconfiguration. Fig. 1 shows a generic SUC creation concept. A single-event process triggers a True Random Generator TRNG leading to select/create randomly one unpredictable and unknown cipher $C_j$ from a large class of a cipher-data-base {$C_1$, $C_2$ … $C_S$} having *S* possible ciphers. For generality, a secret unknown key $Z_i$ may be similarly also created. After this single-event process, all dashed entities in Fig. 1 are then irreversibly deleted and should never be possible to act again.

The resulting cipher is a secret, however unknown and a non-repeatable selection. It is even unknown to the cipher designer himself. Therefore, the designation Secret Unknown Cipher (SUC). Notice that for the functionality of the concept, no need to publish the SUC creation program of the cipher-class which is designated later as the GENIE. In worst case, according to Kerckhoffs's principle, the GENIE is considered as published. Notice that the GENIE is fully seeded by the non-predictable bits of a True Random Number Generator (TRNG).

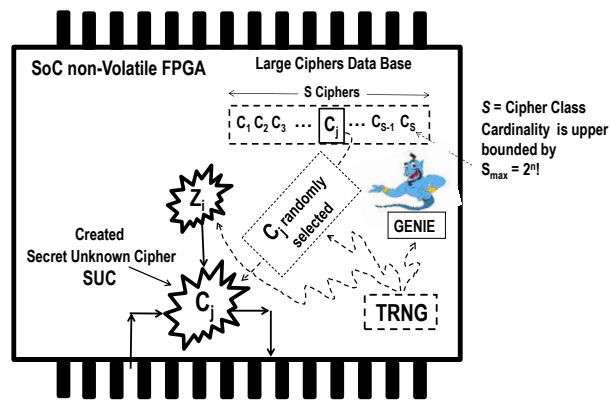

Fig. 1. Key Concept for Creating a Secret Unknown Cipher SUC

*C. SUC Security and Clonability Bounds*

Selecting unpredictable key $Z_i$ from the TRNG output bitstream is basically a trivial process. However, designing unpredictable operational cipher is a very challenging task. The following objectives represent the first obvious challenges and bounds:

1. <u>*SUC-Cloning Resistance:*</u> Basically, something can be cloned if it is known to somebody. The Unclonability of a cipher comes from the fact that nobody knows it. The first obvious challenge that the cipher designer faces is how to create a cipher which the designer himself cannot predict? The first idea is to design a large cipher class from which one unknown cipher is selected. The size *S* of a cipher-class having *n-bit* as input-size is upper bounded by $S_{max}=2^n!$. $S_{max}$ is the number of all possible n-to-n bit permutations including trivial ciphers. Applying Stirling's approximation [14]:

$$S_{max} = 2^n! \approx \left(\frac{2^n}{e}\right)^{2^n}$$

And after further approximations:

$$S_{max} \approx 2^{(n-1.45)\ 2^n} \approx 2^{(n-2)\ 2^n} \quad (1)$$

As a result, cracking/cloning a 64-bit unknown cipher requires a search complexity of $O\left(2^{(64-2)\times 2^{64}} \approx 2^{2^{69.9}}\right)$. Whereas, cloning a standard 64-bit known cipher with unknown secret-key having 64 bits requires to guess just the key with a complexity of $O(2^{64})$. The security difference is obviously tremendous! A created SUC is considered as unknown or highly unpredictable/perfect if it is randomly selected from a class of size $S$ approaching $S_{max}$.

The Cloning Resistance Entropy (CRE) for a SUC can be defined as:

*SUC-Cloning-Resistance-Entropy* $CRE = \log_2 S$

If the creator GENIE is not published then CRE is at maximum:

$CRE_{max} \approx \log_2 S_{max} \approx (n-2)\ 2^n$

If the GENIE is published which is the worst case:

$CRE_{min} = \log_2 S_g$

Where $S_g$ is the class size of the cipher-class offered by the GENIE. The challenge on the cipher designer is to design $S_g$ to be as large as possible with acceptable complexity. The two GENIE created sample cipher classes in this work have the cardinalities $S_{g1} \approx 2^{1350}$ and $S_{g2} \approx 2^{1234}$.

2. *SUC-Modeling Resistance:* A cipher is considered as practically unclonable by modeling if it is not feasible to store all the challenge-response space as the Cipher Codebook size *CCBS*:

$$CCBS = 2^n \geq \psi_0$$

Where $\psi_0$ is the contemporarily valid infeasible cryptographic time and/or memory complexity level. $\psi_0 \approx 2^{80}$ is considered as adequate for traditional contemporary cryptography. $\psi_0 \approx 2^{160}$ is taken as adequate for post-quantum cryptography. For 80-bits ciphers *n=80*, that is *CCBS=$2^{80}$*. For low-cost practical consumer applications even *n=64* may be acceptable.

D. *VLSI Hard-wired One-Way Functions*

The emerging VLSI technology was the driving motivation towards the SUC concept. Self-creating of irreversible, permanent and unknown mappings became first thinkable by the emerging modern non-volatile VLSI technology.

The breakthrough concept of the public-key cryptography introduced in 1976 was based on claimed mathematical one-way functions. All such claimed one-way functions, from which the public-key cryptography still lives are not provably one-way functions and a prove of one-way perfectness may never become possible in the future. Let us define similar functions in VLSI structures.

Assume that the emerging VLSI technology would offer the capability of creating physically one-way hard-wired (permanent) functions in the sense that: they are hard (or even impossible) to reverse, change or remove. We postulate that this would become feasible within future VLSI structures through internal self-reconfiguration. Triggering a device-internal, single-event process without being able to predict/trace the created functions seems to become feasible. In that case; a **one-way physical unknown function** is created. Assume that it is technologically infeasible/hard to reveal and change that created unknown physical digital entity, then an **unclonable physical entity** is created. As a result, the device incorporating that entity becomes unclonable.

In difference to mathematical one-way functions, physical one-way VLSI functions as "encapsulated-secrets" may practically be kept as secret and "unknown mappings" within the devices. Internal one-way-locks as irreversible-locks seem also to be practically realizable as in the case of *anti-fuse* VLSI technology. Recent technologies as *memristors* may emerge in 3D constellations to new interesting permanent VLSI structures having smart one-way physical locking capabilities.

A *pragmatic-security* and practically interesting property of such structures is that, attacking hardwired physical functions require *physical invasive attacks* which may practically-frustrate attackers by being more expensive and occasionally impossible. Moreover, analytical and side channel attacks on unknown locked physical structures is much more complex especially if the physical layout locations of the structures are also not known.

In Summary: Our key-concept assumes that future non-volatile VLSI technologies would become smart-enough to allow self-creating/mutating of permanent unknown secrets or even operational and usable unknown cryptographic functions at unknown physical locations. We further assume that invasive physical attacks on such devices would become practically infeasible especially when creating hard-to-trace physical one-way functions in 3D technologies. The resulting systems may be considered as practically-perfect if invasive attacks do not pay off. This is a very essential aspect for mass products as those in consumer and vehicular technology.

A *new paradigm* like the mathematical one-way functions in public-key cryptography is mapped conceptually into the physical VLSI environment. Physical entities are expected to offer even much larger operational space than mathematical functions as physics is dealing with additional dimensions like space and time compared to the mathematical functions which are limited to abstract mappings.

E. *The Use of Unknown Ciphers and their Advantages*

Traditional PUFs as *unknown weak hash mappings* were used two decades ago [1][2][3] as born DNA-like chains to serve as unclonable identities with all their operational instability issues [15][16]. More details on PUFs are given in section III. PUFs may be seen somehow as a sort of born DNA-like analog weak *unknown hash functions (non-linear functions)*.

There is, to the knowledge of the authors, no PUF-like natural *born unknown cipher mappings* in our physical environment. Artificially created SUCs, as one-to-one *invertible* mappings would offer much larger *full-usability-space* compared with the *non-invertible* PUFs mappings. A generic use-protocol to identify a physical unknown cipher module is described later in section IV.D. Refined SUCs use-protocols were shown to exhibit new applications and superior efficiency and manageability compared with the traditional PUFs as demonstrated in [17][8]. However, the most obvious and practical striking advantage of unknown ciphers compared with PUFs, is their "digital structure" exhibiting ultimate operational consistency over the whole lifetime of the digital VLSI device. In comparison, traditional analog PUFs are highly temperature, operation conditions and aging sensitive leading to very complex realization countermeasures to compensate for such drawbacks. There is a crying need for such consistent, clone-resistant and low-cost physical modules in a large class of contemporary digital mass products. One very important property when deploying SUCs as post-fabrication entities, is that their security is manufacturer-independent as the whole security issues are shifted to the responsibility of the end-user or application manager.

*F. SUC Creation Challenges and Difficulties*

Manufacturers of non-volatile-FPGA like Microsemi stated that the most challenging self-reconfiguration requirement may become technically possible and available soon. However, more challenges would be facing the design and realization of *SUC*s in VLSI environment. The following challenges are identified:

1. How to *design a GENIE* as cipher creating software package and/or cipher database of size $S$ ($S \rightarrow 2^n!$) with practical time and memory complexities to be executed just one time within the SoC unit.

2. *How to place/diffuse the selected cipher* structure at possibly unknown locations in the FPGA-fabric to confuse attackers without violating the design rules of the VLSI structure. This seems to be the most challenging and possibly a very hard to solve task!

3. How to make the *fabric-and memory resources* consumed by the SUC module practically as small as possible.

The design challenges seem at the first sight to frustrate both cipher designers and FPGA programmers. In total, a highly challenging engineering and scientific task.

This work is proposing two SUCs classes with moderate cloning quality ($S_g = 2^{1350}$ or $2^{1234}$) in a realistic environment. Units personalization should operate online at very high speed and relatively low production costs. The diffusion of the ciphers in the floorplan of the FPGA fabric is not perfect as ciphers locations are fixed to allow simple online *SUC*-creation while the bitstream is uploaded into the FPGA devices.

The work represents one of the first attempts toward what we expect to become a VLSI-technology-oriented new paradigm in cryptography.

III. TRADITIONAL PHYSICAL UNCLONABLE FUNCTIONS

Since their advent in the early 2000s [1][2][3], Physical Unclonable Functions (PUFs) were increasingly proposed as central building blocks in security architectures. PUFs are proposed to be used for many applications such as devices identification and authentication [5][6], memoryless key storage [18] and intellectual property protection [19]. Most PUFs response spaces are noisy resulting with limited identity entropy. As a remedy to this problem, fuzzy extractors were proposed to stabilize the output response of each PUF [15][16]. Fuzzy extractors generate and store helper-data during the enrollment phase, which will be used for error correction in the reconstruction phase when the PUF response is noisy to reproduce the correct response by automatic error correction. Such error correction mechanisms are complex, costly and require a large number of logic gates, for instance Intrinsic ID' Quiddikey plus requires 42 K Gates [20]. Compared to PUFs, *SUCs* do not require any error correction mechanism due to their digital structures. Moreover, *SUCs* result with lower hardware complexity and less latency.

Many attacks on PUFs have been recently proposed. They are targeting both weak PUFs and strong PUFs; weak PUFs have few challenges, commonly only one challenge per PUF instance. Hence, it is assumed that access to the weak PUF response is restricted. However, semi-invasive means have been used to reveal the state of memory-based PUFs [21]. The second major PUFs types are Strong PUFs having large number of Challenge-Response Pairs (CRPs) and are assumed to be unpredictable. Hence, protecting the challenge-response interface is not necessary. Strong PUFs are less susceptible to cloning and invasive attacks as weak PUFs. However, modeling attacks represent a strong threat in cloning such PUFs. D. Lim introduced the first attack to model an Arbiter-Based PUF [18] and later on Majzoobi et al. analyzed linear and feed-forward PUF structures [22]. Recently, Rührmair et al. demonstrated PUF modeling attacks on many PUFs by using machine learning techniques [23]. In [24], side channel attack was used to analyze PUFs architecture and fuzzy extractor implementations by deploying power analysis. Recent attack trends combine both side channel and modeling attacks [25][26] to facilitate machine learning deployed for modeling attack. In [27], a hybrid attack is presented, combining side channel analysis and machine learning for attacking especially weak PUFs which prohibit attackers to observe their outputs. The same attack method can also be applied to strong PUFs. It was also shown that fuzzy extractors are also vulnerable to power analysis. This shows in general that traditional PUFs suffer from high-complexity and inherent inconsistency.

IV. THE CONCEPT OF SECRET UNKNOWN CIPHERS AS DIGITAL PHYSICALLY CLONE-RESISTANT FUNCTIONS

*A. Early Work on Secret Unknown Ciphers*

In [28], a first attempt towards *SUC* realization was proposed as a secret and unknown random stream cipher deploying T-functions as key stream generators with random S-Boxes. In [29], a new family of stream ciphers was proposed to be used as *SUCs*, the design is based on combining Nonlinear Feedback Shift Registers (NFSRs) with randomly selected feedback functions from a set of Boolean functions ensuring maximum period NFSRs. Both *SUC* designs are based on random stream ciphers. One of the objectives of this work is to create new efficient and large block cipher classes which can be easily embedded/mutated as *SUCs* in modern SoC FPGA devices. Such *SUCs* may serve as digital-PUFs at adequate cost and time complexities. *SUCs* based on random block ciphers can be efficiently deployed in a wide spectrum of smart vehicular security applications such as security-critical over the air software update [8]. SUC was also proposed as central building block for end-to-end device-to-device authentication in 5G network [9]. Another SUC-based solution for highly secure implantable medical devices was proposed in [10].

*B. Conceptual Definition of SUC*

Secret Unknown Cipher is a randomly internally self-generated cipher inside a chip, where the user has no access or influence on its creation process. Even the device manufacturer should not be able to back-trace the creation process and deduce or predict the created random cipher. An individual unknown cipher is created in each unit after a non-repeatable single event non-reversible process. Each generated *SUC* is an invertible unknown Pseudo Random Function (*PRF*); defined as follows:

$$SUC : \{0,1\}^n \rightarrow \{0,1\}^m \quad (2)$$
$$X \xrightarrow{PRF} Y$$

and its inverse mapping as:

$$SUC^{-1} : \{0,1\}^m \rightarrow \{0,1\}^n \quad (3)$$
$$Y \xrightarrow{PRF^{-1}} X$$

A *SUC* when designed as a block cipher, requires that $m \geq n$. For lowest implementation complexity, an *Involutive-SUC* (*I-SUC*) is a good choice. In that case $SUC = SUC^{-1}$. For optimum space utilization $m = n$, (input and output spaces are equal) the cipher is then defined as a Pseudo Random Involution *PRI*:

$$SUC : \{0,1\}^n \rightarrow \{0,1\}^n \quad (4)$$
$$X \xrightarrow{PRI} Y$$

where $SUC(SUC(X)) = X$ for all $X \in \{0,1\}^n$. That is, encryption and decryption operations use the same mappings resulting with minimized realization complexity.

*C. Basic SUC Creation Concepts and Use Scenario*

Fig. 2 describes a possible scenario for embedding *SUC* in a System on Chip (SoC) non-volatile FPGA devices. The personalization process proceeds as follows:

- **Step 1:** A Trusted Authority (TA) uploads a software package called "GENIE" that contains an algorithm for creating internally secure random ciphers. Possibly, a Cipher Data Base CDB of cryptographically strong functions is included to support selecting the *SUCs*. The TA uploads the GENIE for a short time into each SoC FPGA unit to be used for just one time.
- **Step 2:** After being loaded into the chip, the GENIE is triggered to create a permanent (non-volatile) and unpredictable random cipher. The cipher design components are completely randomly selected by deploying random bits from a True Random Number Generator (TRNG) within the chip.
- **Step 3:** After completing the $SUC_u$ creation, the GENIE is completely deleted
- **Step 4:** by completing step 3, the SoC FPGA unit *u* contains its unique and unpredictable $SUC_u$. TA then personalizes/enrolls the unit *u* by challenging its $SUC_u$ with a plaintext challenge-set $\{X_{u,0}, X_{u,1} \ldots X_{u,(t-1)}\}$ to gets the corresponding ciphertext response-set $\{Y_{u,0}, Y_{u,1} \ldots Y_{u,(t-1)}\}$. The two sets are stored securely as secret records in the Units Individual Records (UIR) labeled by the Serial Number of the device $SN_u$. UIRs are kept secret by TA. A secret key $K_{TA}$ may be added to the *SUC* design for multi TA usage.

The *X/Y* pairs can be used later by TA to identify and authenticate devices.

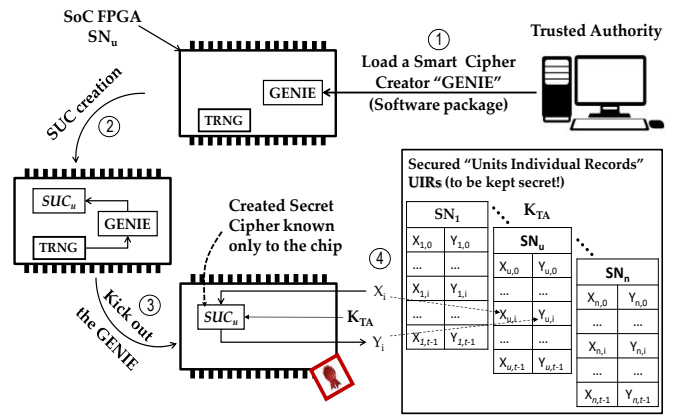

Fig. 2. The basic concept for creating *SUC* in SoC FPGAs environment

*D. Generic Physical Identification Protocol for SUC Units*

After finalizing the personalization process (Step-4 in Fig. 2), TA has for each SoC FPGA a secret record in the UIR including *X/Y* pairs. In reference to Fig. 3, a two-way protocol may identify a physical unit *u* ($SN_u$), having $SUC_u$ and $SUC_u^{-1}$ structures as follows:

- TA selects randomly one of the $X_{u,i}/Y_{u,i}$ pairs and challenges unit *u* with $Y_{u,i}$. Unit *u* uses its $SUC_u^{-1}$ to decrypt $Y_{u,i}$ resulting with the corresponding cleartext $X'_{u,i} = SUC_u^{-1}(Y_{u,i})$ and sends $X'_{u,i}$ to TA.
- If $X_{u,i} = X'_{u,i}$ then the unit is deemed to be authentic and can be accepted. Otherwise *u* is not authentic and should be rejected. The pair $X_{u,i}/Y_{u,i}$ is marked as consumed and should not be used later for highest security performance.

The concept is comparable to a PUF with the advantage that SUC based design is capable to recover *X* from *Y* by using the inverse function SUC$^{-1}$. This property was used in [8] to build a physical chain of trust for a secured over the air vehicular software update.

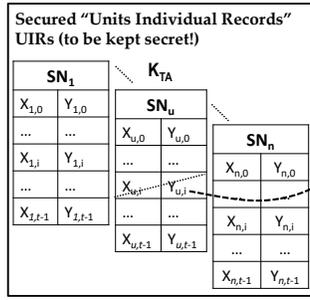
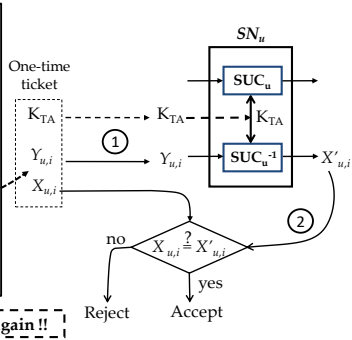

Fig. 3. Generic use protocol for a secret unknown cipher SUC

*E. Requirements for SUC-Creation*

The most difficult and challenging task in creating *SUCs* is how to devise and run a "GENIE" software with acceptable complexity in memory and time in the target SoC units. This is highly dependent on the existing technology infrastructure.

The requirements for an ultimate creation environment can be summarized as follows:

1. A non-volatile FPGA fabric with self-reconfiguration capability, to result with permanent hard-wired structures created by a non-repeatable single-event process by the GENIE program. Such technology is not yet available, but it is expected to be available soon in emerging smarter technologies. The nearest usable existing non-volatile technology is Microsemi SmartFusion®2 SoC FPGA devices. However, without a self-reconfiguration capability.
2. The created *SUC* structures should become non-removable. That is, a secured one-way locking mechanism to prohibit later self-reconfiguration capability is necessary.
3. Low GENIE runtime for cipher creation.
4. Low software complexity and runtime in downloading and deleting the GENIE.
5. The locations of the created *SUC* functions should be hard to find or to predict.
6. The selected *SUC* functions, their parameters and their operation sequence and contents should be unpredictable and analytically hard to attack.
7. The attained *SUC* security-level should be acceptable even when the "GENIE" is completely published, disclosed or somehow become known to the attackers.

V. A NEW CONCEPT FOR *SUCS* CREATION IN FPGA DEVICES

This section presents a novel technique for high-speed creation procedure of *SUCs* in SoC FPGAs. The creation concept is based on Bitstream-manipulation in a pre-defined FPGA layout template.

*A. New Key-Concept for Template-Based SUCs Creation*

The key idea of the creation concept and targets are illustrated in Fig. 4. According to Fig. 4 (a), the design compiler of the FPGA investigates the floorplan of the existing used application area and seeks free gaps of unused FPGA-cells. A designer may also reserve some free layout areas/blocks as free-area to allocate the *SUC* structures there in a later incremental design compilation. The free areas (dashed area in Fig. 4 (a)) are then routed and interconnected as an *SUC-design-template* with default design-rules-safe contents. This *SUC*-design-template is downloaded in the bitstream completely equally for all units to be personalized. A layout for the free templates is shown in Fig. 4 (a). In the personalization process, and when downloading the FPGA-bitstream into each individual unit as in Fig. 4 (b), a random selection of cipher mappings and their parameters is programmed/configured in the free software/hardware templates respectively to make each unit differently unique. The GENIE should create completely differently allocated, and occupied with unpredictable and unknown ciphers as shown in Fig. 4 (b), units 1 to *n*.

Notice that the *SUC*-design-template contains mappings and functionalities which may be or may not be used in the final generated *SUC* inside each individual unit. The software mappings, constituting the *SUC* program, are distributed also randomly in blocks in the reserved non-volatile memory locations. The use and parameter selections of all reserved templates are completely randomly selected by the random bits generated by the TRNG module within the SoC unit. As the TRNG bits are completely unpredictable, the resulting ciphers $SUC_1$ to $SUC_n$ are fully unknown, different and unpredictable in their locations, contents and parameters as symbolically indicated in Fig. 4 (b).

Again, the SoC FPGA configuration bitstream is manipulated by the GENIE affecting only the dedicated *SUC-design-template* locations according to the TRNG random source bits. Therefore, as shown for example in Fig. 4 (b), the unit of $SUC_1$ has less blocks and are differently occupied at different locations compared with the unit of $SUC_n$.

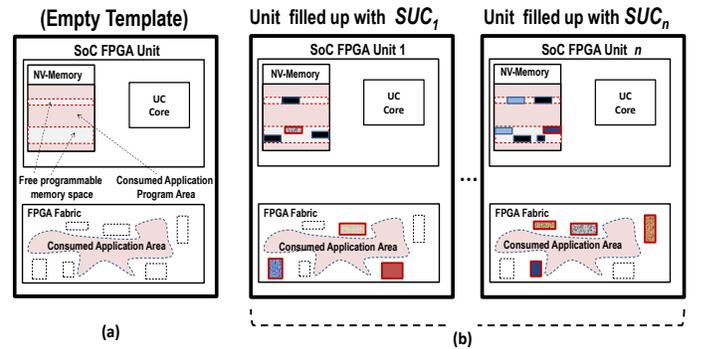

Fig. 4. A Template-based concept for *SUC* generation in FPGA environment

*B. Technical Realization of a Template-Based SUC Creation*

Fig. 5 illustrates a possible internal personalization process. It is also describing a concept for partially self-reconfiguring SoC FPGAs. During the personalization process, the Manipulating GENIE modifies the corresponding bits of the cryptographic mappings by randomly selecting ciphering blocks from the Cipher Data Base (CDB). The blocks are filled up also directly by randomly selected contents from CDB. All random selections are controlled by the TRNG bit source to deliver unpredictable selections. In Fig. 5, the SoC FPGA configuration bitstream *BS* is modified to *BS'* to accommodate the default *SUC*-design-template into it. For FPGA personalization process, *BS'* is usually encrypted to *EBS'* for better production security in the user environment. Fig. 5 illustrates the personalization process that proceeds as follows:

1. The encrypted bitstream *BS' as EBS'* is downloaded equally to configure each SoC FPGA individually. *EBS'* includes the application design together with the *SUC*-design-template.

2. The decryption results with *BS'* bitstream which is manipulated in real time by the GENIE controller fed and instructed by the random bits from TRNG. Based on these true random bits, the controller selects some mappings from the Cipher Data Base (CDB) and possibly selects random fill-ups mappings correspondingly. The control unit manipulates the configuration bitstream $BS'$ to $BS'_u$ for unit *u* correspondingly. The result now is an individual, randomly personalized bitstream $BS'_u$ for the unit *u*. Notice that, even if all units are personalized at the same time in parallel, each unit would create a different unpredictable and unknown own *SUC*. The probability to get two equal *SUC*s, even when key-entropy is ignored, is *1/S*. The attained *SUC-cardinalities* in the two proposed *SUCs* are $S_g = 2^{1350}$ or $S_g = 2^{1234}$ (see section VI and 0) resulting with probability of equal SUCs approaching zero.

3. $BS'_u$ is now stored in both the non-volatile software part and in the FPGA-fabric to permanently-program the device individually. This process represents a single-event *"electronic mutation"* within each SoC-FPGA device.

After completing the personalization of $SUC_u$, the Manipulating GENIE is deleted and a "possible" reconfiguration-lock is irreversibly activated to prohibit any later changes on the FPGA-fabric or the NV-software. The unit would include at this stage a permanent and operational Secret Unknown Cipher $SUC_u$ which nobody knows. This makes an individual FPGA device physically unique with a probability approaching 100%. The only entity which can encrypt and decrypt according to the unknown cipher $SUC_u$ is that individual unit without any possible substitute. Even the trusted authority TA do not know the cipher and cannot fabricate any physically equal unit. Notice also that, the locations of the used templates are not known. An invasive complicated attack is required to read the manipulated bitstream to be able to clone the unit. Each unit needs to be attacked individually, as cloning one unit would not make cloning another unit less complex.

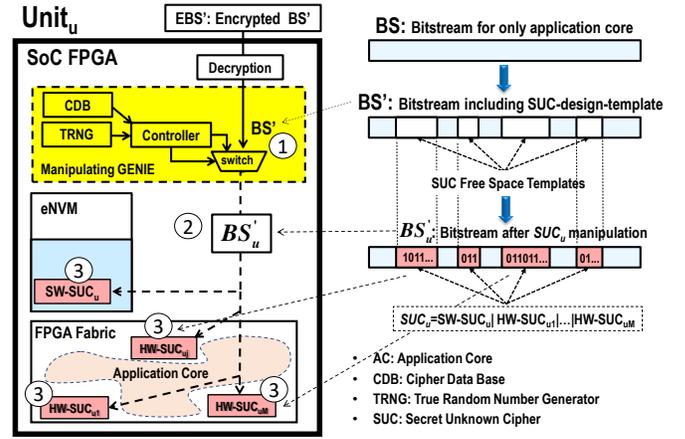

Fig. 5. SoC FPGA personalization/enrollment "mutation" process

### C. Necessary FPGA Infrastructure for SUC Realization

The Configuration bitstream contains all information about all Look Up Tables (LUT) contents, Block SRAMs, MACCs, I/O pins configuration and the exact placement and routing of all the used components in the design. Many modern FPGA technologies encrypt the bitstream to protect designer IPs. A decryption engine in the FPGA device reveals the configuration bitstream internally before configuring the device. In our target NV-Microsemi FPGA technology, even a legal user is not allowed to get access to the clear configuration bitstream. The user has even no information about the bitstream format to allow locating the bits to be manipulated in the *SUC*-design-templates. Therefore, a modified bitstream management in such FPGAs is necessary to allow the realization of the presented SUC creation concept. The modified bitstream management should allow bitstream manipulation and allow knowing the locations of the corresponding footprints of the template in the bitstream format. Some FPGA technologies allow locating LUTs contents in the configuration bitstream. It may also be possible to allocate the configuration bits of the interconnection fabrics inside the FPGA.

Fig. 6 shows a sample 4-to-4-bit S-Box. Each output bit $y_i$ represents a logical function $F_i$ of the inputs $x_3 x_2 x_1 x_0$. Every 4-bit LUT can implement any 4 to 1 logical function. The function $F_i$ is realized in $LUT_i$ through a block of 16-bit in the bitstream describing the truth table of the 4 to 1 logical function. The 4 LUTs required to implement the 4-to-4-bit S-Box are represented by 4 blocks of 16-bits as in the example of Fig. 6. That is, the total configuration bits (bitstream bits) required to implement such 4-bit S-Box is 64 bits.

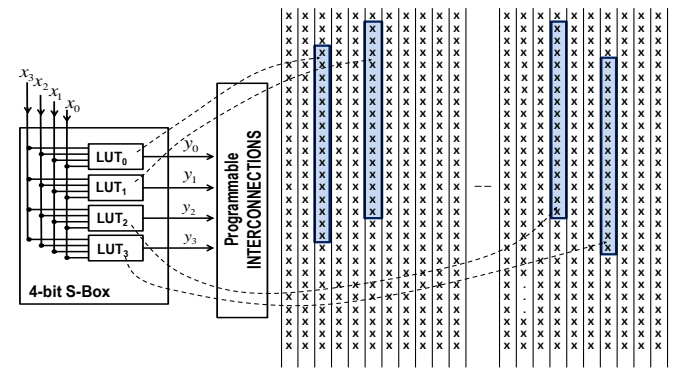

Fig. 6. Sample representation of a 4-bit S-Box in the configuration bitstream

*D. Inspiring SUC Structures from FPGA Resources*

To attain the highest resource utilization the mapping boxes deployed for a cipher should have the same input-size as the LUTs. Consequently, the existing FPGA elements in their most efficient operation and interconnection mode should define and dictate the structure of the mapping operations to be used in the created SUCs. Possible strategies and rules for ultimate creation of *SUCs* in future programmable VLSI technologies can be summarized as follows:

1. Recycle the existing unused free resources in a modern FPGA application device to come up with minimum and possibly zero-cost *SUC* design.
2. Let the existing free logical resources define and dictate the basic cipher mapping functions to be used in designing the GENIE cipher classes.
3. Design new crypto-mappings to optimally use such ready and free existing FPGA functions.
4. The implementation complexity is seen to be zero if the unused free FPGA resources are consumed in the created SUC structure.
5. *SUCs* are accessed rarely in many applications. In such cases, a trade-off between execution time and consumed hardware resources may optimize the whole system performance. The smaller the hardware structure is, the easier is hiding and managing it within the FPGA fabric.
6. Investigate completely new non-conventional ciphering structures based on the free unused FPGA mapping resources. This may result with novel cipher-classes which are otherwise not considered by standard cipher designers due to their unacceptable high-complexity.

*E. Design-Flow for Creating SUCs in Microsemi SoC FPGAs*

To check the feasibility of the proposed SUC creation concept, a sample real FPGA environment is selected as a possible future target technology. SUC concept requires non-volatile FPGA devices to make sure that the created SUCs as units' identities are permanent and not removable. Modern SoC FPGAs incorporate an FPGA fabric together with at least a microcontroller core. Microsemi FPGA technology is to our knowledge, the only contemporary non-volatile FPGA technology having the non-volatile configuration memory cells distributed over the whole NV-FPGA fabric area. This makes an invasive attack more difficult if the bitstream is not readable as the whole fabric area need to be attacked. In contrary to that, volatile RAM-based FPGAs, can be attacked when reloading the bitstream in the start-up after every power-off case. That is, the attack can focus on the flash memory accommodating the bitstream and possibly its usually deployed encryption mechanism. Non-volatile technology does not require to reload bitstreams after power-off offering much higher physical security level. Repeated transfer of secrets is a serious security threat in all practical systems. Volatile FPGAs can only be securely used to accommodate SUC modules if the operational lifetime of the unit's identity is just the time between a single switch-on and off phase. This represents a dramatic limitation for most physical security applications.

Fig. 7 describes a possible design flow and device personalization/enrollment process for creating *SUCs* in SoC FPGAs. The sketched concept is dedicated for Microsemi SoC FPGAs architecture and may be applicable on similar FPGA technologies. The process may proceed as follows:

1. The *SUC* design template structure is added to the dedicated device Application Core (AC) that should be compiled and locked against layout changes in an early stage.
2. If no SUC creation is necessary, Libero SoC creates a pure application bitstream *EBS* corresponding to device AC.
3. To generate a bitstream including the *SUC*-design-template, Libero SoC can perform an incremental synthesis and routing for the free *SUC*-design-templates. The updated Encrypted Bitstream (*EBS'*) contains the application core together with the *SUC*-design-template.
4. When loading *EBS'* into the device, a decryption engine residing inside the FPGA decrypts *EBS'* to result with a clear bitstream *BS'*.
5. The Manipulator GENIE residing in each SoC modifies the bitstream *BS'* on-the-fly during its loading by filling up the free templates by the randomly chosen *SUC* mappings in every particular device. The manipulator GENIE delivers then the personalized bitstream, for instance $BS_1'$ for unit 1.
6. After personalization, the Manipulator GENIE is deleted. Each SoC unit includes its own mutated *SUC* identity.

Steps 4 to 6 are repeated for each unit individually.

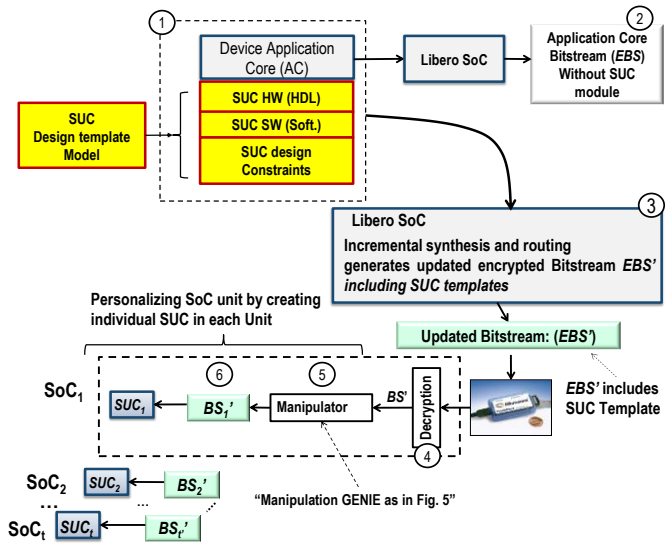

Fig. 7. Possible process flow for creating *SUCs* in Microsemi-similar SoC FPGA environment

Fig. 8 describes in details a possible schedule to compile the new bitstream *EBS'* including the free *SUC* bitstream-templates embedded in the device application core. The process can proceed as follows (Fig. 8):

1. The user application core is locked after compilation. Its layout can also be locked to avoid any influence on the original application tasks and performance.

2. The *SUC* design template is added in HDL to the locked Application Core.
3. The FPGA design-compiler routes incrementally the designed *SUC*- Templates.
4. A new encrypted bitstream *EBS'* including the application core together with the free SUC templates is created.

The same *EBS'* should be generated in encrypted form for each device to be programmed/enrolled.

***Necessary FPGA Changes:*** *Notice that the Microsemi-FPGA bitstream management requires to be modified to allow embedding the bitstream manipulator GENIE shown in Fig. 5 and deleting it after use. In addition to that, the design compiler should allow the cipher-designer to know the locations of the bits corresponding to the created free templates.*

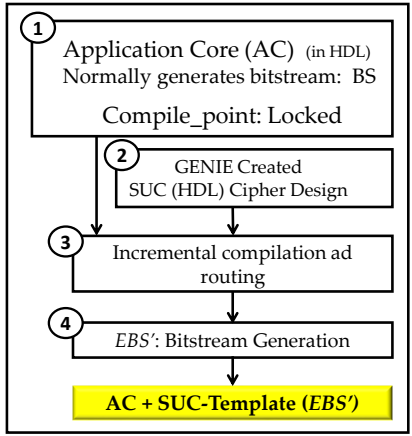

Fig. 8. Bitstream incremental update for embedding the *SUC*-design-templates

The SUC-templates are generated according to the GENIE-created cipher class. Two sample SUC cipher-classes are proposed in the following sections.

## VI. A NON-INVOLUTIVE *SUC* STRUCTURE PROPOSAL

For efficient performance, *SUC* should consume low software and hardware FPGA resources to be implemented possibly at zero cost in the free available FPGA resources. This section describes a first proposed *SUC* class coined as a Non-Involutive *SUC* (*NI-SUC*).

### A. A New Concept for Non-Involutive SUC Design Structure

The proposed non-Involutive SUC (*NI-SUC*) is a random cipher based on Substitution Permutation Network (SPN) structure designed to optimally use the FPGA resources. Fig. 9 describes the overall *NI-SUC* cipher structure having a block size of 64 bits. The cipher includes *R*=31 rounds where $R+1$ keys are generated by a key scheduling algorithm. For each round, a round key $K_i$ is XORed before the substitution layer and after the last round.
The substitution layer includes 16 randomly 4-to-4-bit mappings selected from the list of all optimal 4-bit S-Boxes. Optimal 4-bit S-Boxes were designed and investigated in [30] to be optimal against differential, linear and algebraic attacks. The diffusion stage is implemented as fixed bit permutation.

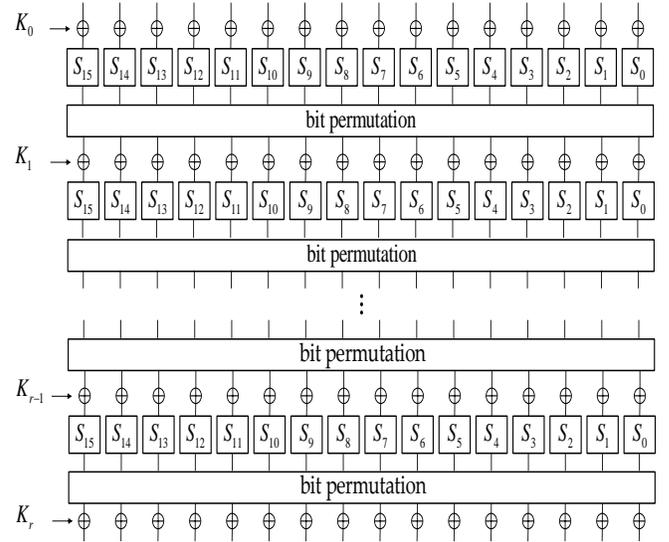

Fig. 9. *NI-SUC* cipher structure

### B. Optimal 4-bit S-Boxes for the Creation GENIE Library

Let $S : \mathbb{F}_2^4 \rightarrow \mathbb{F}_2^4$ be a 4-bit S-Box *S*. Let $Lin(S)$ and $Diff(S)$ denote the linearity and the differential resistance of *S* respectively. An optimal 4-bit S-Box fulfills the following conditions [30][31]:

1. *S* is a bijection
2. $Lin(S) = 8$
3. $Diff(S) = 4$

In [30], the affine equivalence classes of bijective 4-bit S-Boxes have been exhaustively analyzed. There exist $1396032 \approx 2^{20.4}$ such optimal 4-bit S-Boxes ordered in 16 classes.
The proposed substitution layer of *NI-SUC* contains 16 randomly selected optimal 4-bit S-Boxes out of the set of all $1396032 \approx 2^{20.4}$ optimal 4-bit S-Boxes.

### C. Non-Involutive SUC Bit Permutation Mapping

To attain adequate hardware efficiency, we propose to use a bit permutation stage which deploys only the interconnections fabric for the diffusion layer. A new fixed bit-permutation is devised and investigated to be deployed in the *NI-SUC* design.

TABLE I
*NI-SUC* BIT-PERMUTATION

| *i* | 0 | 1 | 2 | 3 | 4 | 5 | 6 | 7 |
|---|---|---|---|---|---|---|---|---|
| *p(i)* | 0 | 4 | 8 | 12 | 16 | 20 | 24 | 28 |
| *i* | 8 | 9 | 10 | 11 | 12 | 13 | 14 | 15 |
| *p(i)* | 32 | 36 | 40 | 44 | 48 | 52 | 56 | 60 |
| *i* | 16 | 17 | 18 | 19 | 20 | 21 | 22 | 23 |
| *p(i)* | 1 | 5 | 9 | 13 | 17 | 21 | 25 | 29 |
| *i* | 24 | 25 | 26 | 27 | 28 | 29 | 30 | 31 |
| *p(i)* | 33 | 37 | 41 | 45 | 49 | 53 | 57 | 61 |
| *i* | 32 | 33 | 34 | 35 | 36 | 37 | 38 | 39 |
| *p(i)* | 2 | 6 | 10 | 14 | 18 | 22 | 26 | 30 |
| *i* | 40 | 41 | 42 | 43 | 44 | 45 | 46 | 47 |
| *p(i)* | 34 | 38 | 42 | 46 | 50 | 54 | 58 | 62 |
| *i* | 48 | 49 | 50 | 51 | 52 | 53 | 54 | 55 |
| *p(i)* | 3 | 7 | 11 | 15 | 19 | 23 | 27 | 31 |
| *i* | 56 | 57 | 58 | 59 | 60 | 61 | 62 | 63 |
| *p(i)* | 35 | 39 | 43 | 47 | 51 | 55 | 59 | 63 |

TABLE I describes the permutation table. The bit position $i$ of round $r$ is mapped to bit position $p(i)$ of the next round $r+1$. This bit permutation has the property that the outputs of any S-Box in round $r$ are connected to the inputs of 4 different S-Boxes in round $r+1$.

### D. Proposed Key Scheduling Algorithm

The LUTs in a non-volatile technology can be deployed to efficiently store keys. Fig. 10 shows a novel Random Key Scheduling Algorithm for *NI-SUC* ( $RKSA_{NI}$ ), to accommodate 32 fully random keys of length 64 bits in 64 LUTs. It is based on using random 4 to 1 mappings $F_{k_i^j} : C \rightarrow k_i^j$, where $C$ is the 4 bits LSB or the 4 bits MSB of the rounds counter and $k_i^j$ denotes the subkey bit $j$ of round key $i$. All the mappings $F_{k_i^j}$ are 4-to-1 LUTs. Hence one LUT stores 16 key bits. The key scheduling requires 64 LUTs for all random mappings $F_{k_i^j}$.

To realize $RKSA_{NI}$, the GENIE manipulates randomly the configuration bitstream for the 64 LUTs by direct use of the random bits generated by the TRNG. The whole created key is then fully unpredictable and unknown. The up-down counter is used to allow both encryption and decryption operations. $RKSA_{NI}$ generates the encryption round-keys by counting-up and the decryption round-keys by counting-down.

As each LUT is a 4 to 1 mapping having $2^{2^4} = 2^{16}$ possibilities. Hence the cardinality (key space) of $RKSA_{NI}$ is:

$$|RKSA_{NI}| \approx 2^{2^4 \times 64} = 2^{1024}$$
$$\text{Or the key} - Entropy = 1024\ Bits \qquad (5)$$

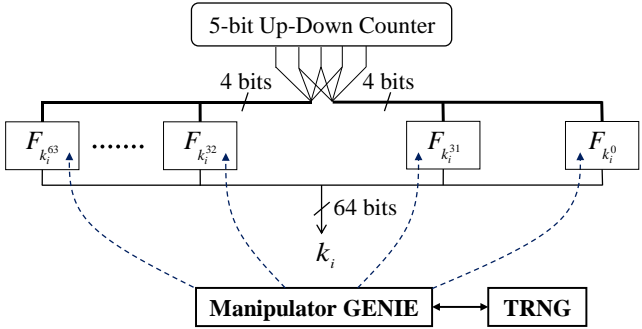

Fig. 10. Random key scheduling algorithm for *NI-SUC* ( $RKSA_{NI}$ )

### E. The Cardinality of NI-SUC Cipher Class

*NI-SUC*s deploy 16 randomly selected 4-bits S-Boxes from the set of all optimal 4-bits S-Boxes and a fixed bit permutation. Each resulting *SUC* depends on the randomly selected S-Boxes and the bits of the key scheduling algorithm.

For *NI-SUC*, with a block size of *N-bits* and number of different substitution layers |*SL*|, using 4-bit optimal S-Boxes, the cipher class cardinality with r-different rounds is:

$$|NI\text{-}SUC| = 2^{\log_2(\#\text{Optimal 4-bit S-Boxes}) \times |SL| \times \frac{N}{4}} \qquad (6)$$

For a block size of $N = 64\ bits$ and one distinct round |*SL*|=1, the number of all creatable ciphers (cipher-class cardinality) is:

$$|NI\text{-}SUC| = 2^{\log_2(2^{20.4}) \times \frac{64}{4}} \approx 2^{326}$$

As the key space is $|RKSA_{NI}| \approx 2^{1024}$. That is $2^{1024}$ different key choices are possible for each *NI-SUC* S-Boxes set selections. Therefore, the total number of possible creatable SUCs with secret unknown keys is:

$$S_{g1} = |NI-SUC|_{RKSA_{NI}} \approx 2^{1024+326} \approx 2^{1350}$$

A single unknown choice out of this huge number of ciphers can be considered as a practically unknown cipher with a cloning resistance entropy of 1350 bits. It is upper bounded by $\approx 20.4 \times 31 \times 16 + 1024 = 11\ 142$ bits for 31 rounds.

## VII. INVOLUTIVE SUC CREATION CONCEPT

### A. Involutive SUC Structure

This section describes an involutive SUC coined as *I-SUC*. The ciphers structure is also inspired from the existing FPGA fabric resources. The cipher is also an SPN design structure with 32-rounds (*R*=32).

Fig. 11 describes the proposed *I-SUC* block structure. *I-SUC* uses in difference to the former cipher class 16 randomly selected however, only Involutive 4-bit S-Boxes ($IS_0$ to $IS_{15}$). Moreover, *I-SUC* deploys a fixed involutive diffusion layer. Notice that the final round includes only a substitution layer.

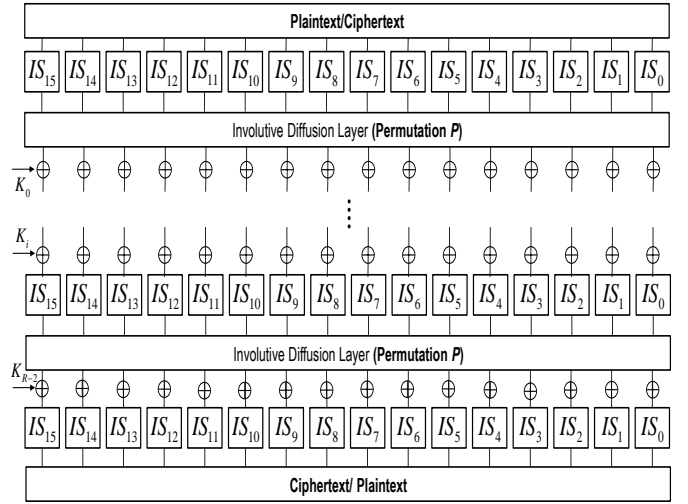

Fig. 11. *I-SUC* using micro involutive S-Boxes

### B. A Library for Involutive S-Boxes

In [31], all the 4-bit involutive (self-inverse) S-boxes with linear, differential and almost resilient analysis were investigated. According to [31], the number of such optimal involutive 4-bit S-Boxes is found to be $145\ 920 \approx 2^{17.15}$

### C. Involutive Diffusion Layer

To use the same structure for both encryption and decryption operations, the diffusion layer should also be an involution. An involutive linear transformation from [32] is modified for the proposed cipher as shown in Fig. 12 and described formally in equations (7) and (8). It is selected and adapted to optimally make use of the hosting FPGA resources.

The linear transformation is only involutive for even number of S-Boxes. Let $S_i^{out}$ be the 4-bit output of the $i^{th}$ S-Box, the XOR sum of the outputs of all 16 involutive S-Boxes is:

$$Sum = \bigoplus_{i=0}^{15} S_i^{out} \quad (7)$$

Each 4-bits output ($Out_i$) of the permutation is defined as:

$$Out_i = S_i^{out} \oplus Sum \quad (8)$$

The overall involutive diffusion layer $P$ is defined as:

$$P(X = S_{15}^{out} \| S_{14}^{out} \| ... \| S_0^{out}) = Out_{15} \| Out_{14} \| ... \| Out_0$$

Where $\|$ denotes the concatenation operation.

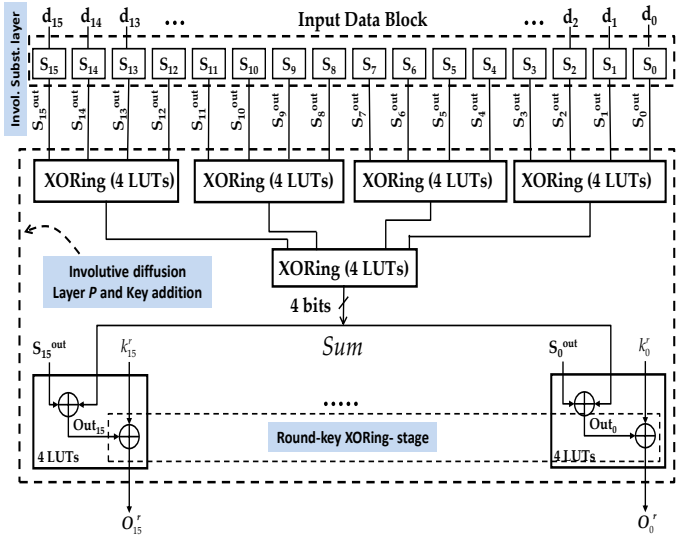

Fig. 12. *I-SUC* round having involutive substitution and diffusion stages with an XORed round key

### D. Conditions on I-SUC to Yield an Involutive Cipher

For $R$ rounds *I-SUC*, let $SL_i$ with $0 \leq i \leq R-1$ be a random involutive Substitution Layer $i$, and the fixed diffusion layer $P$. Let $X$ be the input plaintext and $Y$ be the output ciphertext, then:

$$Y = I\text{-}SUC(X) \quad (9)$$
$$= SL_{R-1}\left(K^{R-2} \oplus P\left(SL_{R-2}\left(...\left(K^0 \oplus P(SL_0(X))\right)\right)\right)\right)$$

For an *I-SUC* to be involutive, the decryption function should be the same as the encryption function however with reversed key order, that is:

$$X = I\text{-}SUC(Y) \quad (10)$$
$$= SL_{R-1}\left(K^0 \oplus P\left(SL_{R-2}\left(...\left(K^{R-2} \oplus P(SL_0(X))\right)\right)\right)\right)$$

If $P$ and key XORing operations commute, such that $P(X \oplus K^r) = P(X) \oplus K^r$ in any round $r$ (see the necessary conditions in the theorem and key deriving equation (14) below) then :

$$X = SL_{R-1}\left(P\left(K^0 \oplus SL_{R-2}\left(...\left(P\left(K^{R-2} \oplus SL_0(Y)\right)\right)\right)\right)\right) \quad (11)$$

In that case the same hardware mappings can be used for encryption and decryption by just reversing the key order.

To fulfill (11), and let *I-SUC* be involutive, the substitution layers need to fulfill the following conditions:

$$SL_{R-i-1} = SL_i \quad \text{with} \quad i \in \begin{cases} \left[0 : \frac{R}{2}-1\right] & ; \text{for } R \text{ even} \\ \left[0 : \frac{R-1}{2}-1\right] & ; \text{for } R \text{ odd} \end{cases} \quad (12)$$

Designing *I-SUC* according to (12) results with a large class of *I-SUCs* since a new set of the involutive S-Boxes can be randomly selected in about 50% of the cipher rounds. However, this would result with high hardware complexity. Hence, we propose to use the same random substitution layer $SL_0$ equally in all rounds.

To fulfill the required Commutative Property between $P$ and the XORed Key, the following theorem followed by deriving design requirements.

**Theorem.** Let $P$ denotes the diffusion layer of *I-SUC*, and let the key of round $r$ be $K^r = K_{15}^r \| K_{14}^r \| ... \| K_0^r$.
If $K_{15}^r \oplus K_{14}^r \oplus ... \oplus K_1^r \oplus K_0^r = 0$, then:

$$P(X \oplus K^r) = P(X) \oplus K^r \quad (13)$$

**Proof.** Let $O^r = O_{15}^r \| O_{14}^r \| ... \| O_0^r$ be the output result of the diffusion layer output XORed with the round key $K^r$, such as:

$$P(X = S_{15}^{out} \| S_{14}^{out} \| ... \| S_0^{out}) \oplus K^r = O^r$$

According to (7) and (8), we have:

$$Sum = \bigoplus_{i=0}^{15} S_i^{out} = S_0^{out} \oplus S_1^{out} \oplus ... \oplus S_{15}^{out}$$
$$O_i^r = Out_i \oplus K_i^r = Sum \oplus (S_i^{out} \oplus K_i^r)$$

To compute $P(X \oplus K^r) = O^{r'}$, first we have:

$$Sum' = \bigoplus_{i=0}^{15} (S_i^{out} \oplus K_i) = (S_0^{out} \oplus ... \oplus S_{15}^{out}) \oplus (K_{15}^r \oplus ... \oplus K_0^r)$$
$$O_i^{r'} = Sum' \oplus (S_i^{out} \oplus K_i^r)$$

For $O_i^{r'} = O_i^r$, the following condition should hold:

$$Sum = Sum' \Rightarrow K_{15}^r \oplus ... \oplus K_0^r = 0$$

**Key requirements to fulfill** (13)**:** This condition can be fulfilled by taking any 4-bit key symbol equal to the XORed sum of the remaining key symbols, for instance:

$$K_{15}^r \oplus ... \oplus K_1^r = K_0^r \quad (14)$$

### E. Key Scheduling Algorithm for I-SUC

Fig. 13 describes the Random Key Scheduling Algorithm for *I-SUC* design ($RKSA_I$) for $R=32$ rounds with 31 keys. In each round $r$, a round key $K^r = K^r_{15} \| K^r_{14} \| ... \| K^r_0$ is generated according to (14) such that $K^r_{15} \oplus ... \oplus K^r_1 = K^r_0$. In this sample design, 60 LUTs are filled-up as unknown secret key randomly by the GENIE while 20 LUTs implement the XORing operations to generate $K^r_0$. In this case, the key cardinality is:

$$|RKSA_I| \approx 2^{2^4 \times 60} = 2^{960} \qquad (15)$$

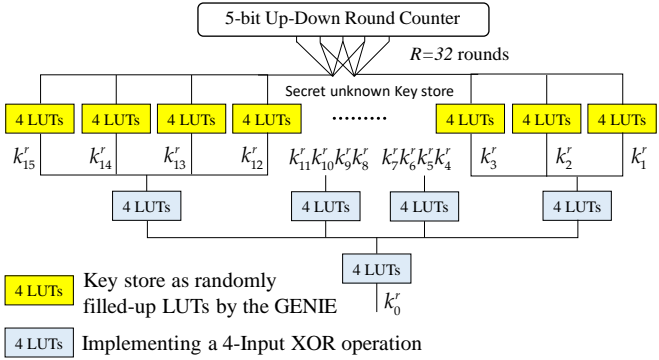

Fig. 13. Key scheduling algorithm for *I-SUC*

### F. Total Cardinality of the Created I-SUC Classes

*I-SUC* deploys 16 random S-Boxes selected from the class of optimal 4-bit involutive S-Boxes, and a fixed linear diffusion layer. For *I-SUC*, with a block size $N$, using optimal 4-bit involutive S-Boxes, the cipher-class cardinality due to the number of possible S-Boxes choices for $|SL|$ different rounds is:

$$|I\text{-}SUC| = 2^{\log_2(\#\text{Optimal inv. 4-bit S-Boxes}) \times \frac{|SL|}{2} \times \frac{N}{4}} \qquad (16)$$

For *I-SUC* with a block size $N=64$ bits, $|SL|=1$, the class-size is:

$$|I\text{-}SUC| = 2^{\log_2(145920) \times \frac{N}{4}} \approx 2^{274}$$

Notice that the key space is $|RKSA_I| \approx 2^{960}$. That is $2^{960}$ different key choices do exist for each *I-SUC* S-Boxes choice. Hence, the total number of possible *I-SUCs* with keys is:

$$S_{g2} = |NI-SUC|_{RKSA_I} \approx 2^{960+274} \approx 2^{1234}$$

*I-SUC* is a practical low-cost SUC version with a cloning-resistance-entropy of 1234 bits. It is upper bounded by $\approx 17.15 \times (32/2) \times 16 + 960 = 5\,350$ bit for 32 different rounds.

## VIII. Security Analysis of the proposed SUC Classes

The proposed cipher classes are ciphers with unknown S-Boxes and keys. Therefore, first the traditional known attacks on ciphers with similar structures are presented.

At Eurocrypt 2001, Biryukov and Shamir investigated the security of AES-like ciphers with key dependent S-Boxes and affine transformation, they successfully cryptanalyzed two and half rounds [33]. The attack was improved in [34] to successfully cryptanalyze Substitution Affine Networks (SANs) of three rounds with unknown or key dependent *m*-bit S-boxes and $n \times n$ bits affine layers with complexities $2^{2m}$ chosen plaintexts and $2^{3m}(n/m)$ time steps.

In [35], Borghoff et al. proposed several attacks on C2 algorithm, which has a secret 8 to 8 bits S-Box and a 56 bits key. The attack reverses firstly the S-Box with complexity of $2^{24}$ C2 encryptions and then the key with a complexity of $2^{48}$. In [36], Borghoff et al. deployed similar manners with differential-style attack to break Maya with a number of rounds up to 28. Maya [37] is a PRESENT-like cipher with key dependent S-Boxes. The attack model described in [36] suggests that it is possible to break up to 28 rounds before reaching the bound $2^{64}$ of possible plaintexts. The attack was extended to break PRESENT-like ciphers with secret components (S-Boxes or bit permutation), and randomly chosen components up to 16 rounds with data complexity less than $2^{64}$. This type of differential-style attack exploits the weak differential properties of key dependent S-Boxes or randomly selected S-Boxes. We propose to deploy only a set of optimal S-Boxes characterized by strong differential and linear properties to prohibit the differential-style attack of [36] for both *NI-SUC* or *I-SUC*.

To cryptanalyze *NI-SUC*, an adversary should reverse the randomly selected optimal S-Boxes and apply a known attack such as differential/linear cryptanalysis to break one *NI-SUC*.

In the following, for both linear and differential cryptanalysis, only the second part of the attack on *NI-SUC* and *I-SUC* are investigated.

### A. Linear Cryptanalysis

This section presents the security analysis of the proposed random block ciphers against linear cryptanalysis [38].

**Lemma 1.** For *NI-SUC*, by deploying the bit permutation in Table I, the number of S-Boxes involved in any 2-Rounds of a linear approximation is greater than or equal to 4.

**Lemma 2.** For *I-SUC*, by deploying the involutive diffusion layer of Fig. 12, the number of involutive S-Boxes involved in any 2-Rounds of a linear approximation is greater than or equal to 2.

The maximal bias of a linear approximation of an optimal *n*-bit S-Box ($n=4$) is:

$$P_\varepsilon = \frac{2^{n-1} - NL}{2^n} = 2^{-2}$$

Where the nonlinearity of optimal 4-bit S-Boxes is $NL=4$.

The number of Plaintext/Ciphertext pairs $N_L$ required for a linear cryptanalysis of an $R$ rounds *NI-SUC* or *I-SUC* is:

$$N_L \geq \frac{1}{P_\varepsilon^{2R}} = 2^{4R} \qquad (17)$$

For *NI-SUC* or *I-SUC* having over 30 rounds, $N_L$ is greater than or at least $2^{120}$, which fulfills the contemporary security level requirement.

## B. Differential Cryptanalysis

Differential cryptanalysis was firstly introduced by Biham and Shamir in [39]. It has been used to break many ciphers such as the full 16 round DES-like cipher [40]. It exploits the high probability of certain occurrences of plaintext differences and differences into the last round of the cipher.

**Lemma 3.** For the *NI-SUC*, by deploying the bit permutation in Table I, the number of S-Boxes involved in any 2-Rounds of a differential characteristic is greater than or equal to 4.

***Proof.*** Optimal 4-bit S-Boxes have the characteristic that, changing a single input bit would change at least two output bits. The proposed bit-permutation in Table I interconnects such that the input bits of any S-Box come from 4 distinct S-Boxes or equivalently any 4 output bits of an S-Box go to 4 distinct S-Boxes.

**Lemma 4.** For the *I-SUC*, after mapping through the involutive diffusion layer in Fig. 12, the number of involutive S-Boxes involved in any 2-Rounds of a differential characteristic is greater than or equal to 4.

The maximum XOR pair probability of the S-Boxes used in the *NI-SUC* or *I-SUC* is equal to $2^{-2}$. Hence, the number of plaintext/ciphertext pairs required for differential cryptanalysis of an *R* rounds *NI-SUC* or *I-SUC* can be approximated by:

$$N_D \geq 2^{4R} \quad (18)$$

For *NI-SUC* or *I-SUC* having at least 30 rounds, the number of Plaintext/Ciphertext pairs required for differential cryptanalysis is greater than or equal to $2^{120}$.

## C. Post-Quantum Cryptanalysis

In 1994, Shor developed new quantum polynomial-time algorithms for the discrete logarithm and integer factoring problems [41]. Shor's algorithm can be used, by an adversary armed with a quantum computer, to break the widely used RSA cryptosystem, DSA and ECDSA. However, there are many classes of cryptography that are beyond RSA, DSA and ECDSA that are not vulnerable to Shor's algorithm such as [42]: Hash-based cryptography, Code-based cryptography, Lattice-based cryptography, Multivariate-quadratic-equations cryptography and Secret-key cryptography. In [43], Grover proposed a quantum algorithm that can find an element in a set of *N* completely randomly ordered elements with a complexity $O(\sqrt{N})$. Grover's algorithm is the only known quantum algorithm threatening symmetric cryptography [44]. It is not shockingly fast as Shor's algorithm, for instance, AES-128/AES256 provide a post-quantum security level of 64-bit/128-bit. For *I-SUC* and *NI-SUC*, the effective attack-complexity is $2^{274}$ and $2^{326}$ respectively (without considering the cardinality of the key scheduling for *NI-SUC*). That is, the attack complexity of Grover's algorithm is $O(2^{137})$ for *I-SUC* and $O(2^{163})$ for *NI-SUC*.

## D. Statistical Analysis of I-SUC and NI-SUC Classes

This section shows some statistical cipher-performance figures for both proposed designs: *I-SUC* and *NI-SUC*.

*1) Simulated Avalanche Behavior of I-SUC and NI-SUC*

Few experimental simulations were conducted to investigate the effect of the number of rounds on the avalanche characteristics of both *I-SUC* and *NI-SUC* designs.

One thousand randomly generated inputs were used to measure the avalanche characteristics for both *SUC*s as a function of the number of rounds. Fig. 14 and Fig. 15 show the experimental results for the number of output bit changes as a function of the number of rounds for both *I-SUC* and *NI-SUC* respectively. The results show that *I-SUC* reaches a perfect avalanche characteristic after only 3 rounds, while *NI-SUC* requires 7 rounds. This is due to the linear transformation which significantly improves the avalanche characteristic of *I-SUC*.

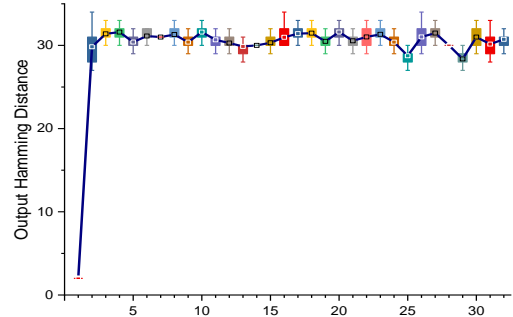

Fig. 14. Output hamming distance as a function of number of rounds in *I-SUC*

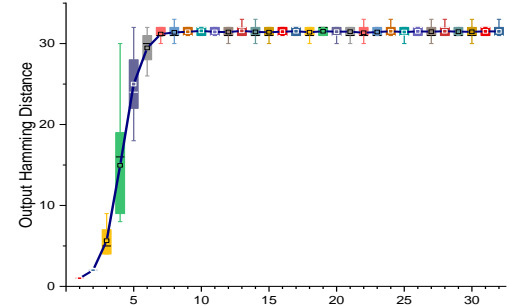

Fig. 15. Output hamming distance as a function of number of rounds in *NI-SUC*

*2) Evaluating the Avalanche Properties of the Cipher Classes*

*SUC* is considered to have a good avalanche characteristic if flipping one input bit results with changing about half of the output bits. Since there are large number of possible ciphers for both proposed *I-SUC* and *NI-SUC* designs, we evaluate the avalanche characteristics for sets of randomly selected *SUC*s as follows:

- One thousand optimal 4-bit S-Boxes are selected to generate thousand *NI-SUC*s
- One thousand involutive optimal 4-bit S-Boxes are selected to generate thousand *I-SUC*s

To evaluate the avalanche characteristics for each resulting *SUC*, 100 random messages are used, where each time one bit of the message is flipped.

Fig. 16 and Fig. 17 show the ranges of the measured number of output bit changes when changing one input bit for thousand *I-SUC*s and *NI-SUC*s respectively. The *SUC*s are labeled as S0, S1, …, S999. Fig. 16 shows that for all randomly selected *I-*

*SUC*s, the expected number of output bit changes is bounded between 28 and 35. Whereas, it is bounded by 22 and 31 for *NI-SUC* in Fig. 17.

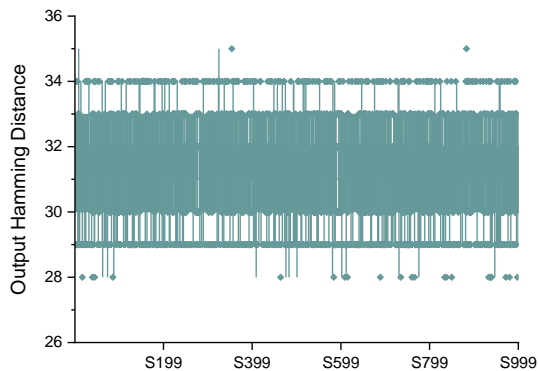

Fig. 16. Avalanche characteristic of one thousand *I-SUC* units

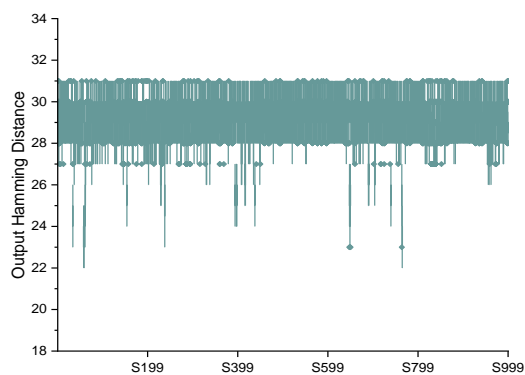

Fig. 17. Avalanche characteristic of one thousand *NI-SUC* units

## IX. PERFORMANCE AND COMPLEXITIES

### A. Hardware Complexity of the Created SUCs

Hardware complexity was one of the main objectives of this work. *NI-SUC* uses a bit permutation mapping that does not consume logic resources (and no additional area in our example). However, only the encryption operation is prototyped and when needed (such as for identification protocol in Fig. 3), the decryption algorithm should be additionally implemented. The decryption module requires about 64-LUT less than the encryption one because both encryption and decryption designs use the same key scheduling. *I-SUC* is an involution and hence both encryption and decryption operations can be performed by using the same structures. When designing the proposed *SUC*s, the highest resource efficiency was targeted to optimally exploit the provided resources in Microsemi FPGAs. The proposed random *SUC*s are lightweight and could be used as physical identities adding a security value (possibly for free) in existing SoC FPGA applications. Fig. 18 shows an area optimized implementation method for the proposed *NI-SUC*. *I-SUC* can be implemented similarly by using the 64-bit register after the substitution layer since the last round does not include a diffusion stage.

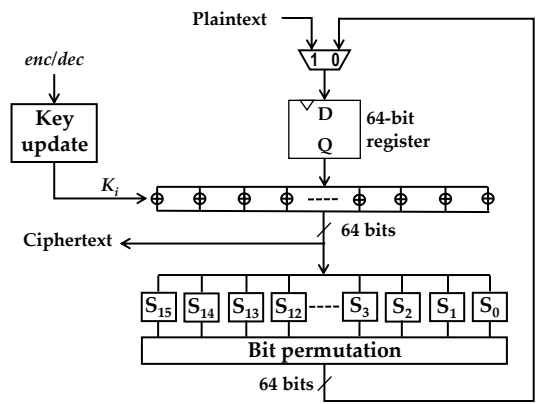

Fig. 18. Area optimized implementation for *NI-SUC*

Table II shows the prototyped hardware complexities of the proposed SUC modules in SmartFusion®2 SoC FPGAs. It shows that the *I-SUC* version is a more efficient design (as involutive cipher) requiring much less resources for both encryption and decryption compared with the *NI-SUC* version.

TABLE II
HARDWARE COMPLEXITIES OF *NI-SUC* AND *I-SUC*
IN SMARTFUSION®2 SOC FPGA TECHNOLOGY

| SUC Design | Resources | Area | Usage in % for M2S05 | Usage in % for M2S150 |
|---|---|---|---|---|
| NI-SUC | LUT | 212 | 3.49 | 0.14 |
|  | DFF | 72 | 1.18 | 0.05 |
| I-SUC | LUT | 226 | 3.56 | 0.14 |
|  | DFF | 72 | 1.18 | 0.05 |

### B. Complexity of a Pure Software SUC

The implementation of *NI-SUC* and *I-SUC* in pure software form is prototyped on the ARM Cortex-M3 core embedded in SmartFusion®2 SoC FPGA. Table III describes the overhead of the software complexities for both *I-SUC* and *NI-SUC*.
Note that M2S010/25/50/60 embed 64 Kbytes of eSRAM and 256 Kbytes of eNVM. Whereas M2S090 and M2S150 embed 64 Kbytes of eSRAM and 512 Kbytes of eNVM.

TABLE III
SUC TEMPLATES SOFTWARE COMPLEXITIES

| SUC Design | Code size Overhead (Bytes) | | FPGA Memory Utilization (%) for M2S0xx | | | |
|---|---|---|---|---|---|---|
| | | | 10/25/50/60 | | 90/150 | |
| | eNVM | eSRAM | eNVM | eSRAM | eNVM | eSRAM |
| NI-SUC | 1152 | 272 | 0.43 | 0.41 | 0.21 | 0.41 |
| I-SUC | 896 | 272 | 0.34 | 0.41 | 0.17 | 0.41 |

## X. SOFTWARE HARDWARE AND TIME COMPLEXITIES OF THE GENIE

The most challenging task when dealing with the *SUC* concept is in designing an efficient GENIE program. The runtime complexity is one of the most challenging tasks for industrial applications. The necessary memory for the GENIE's program and data should be accommodated completely within the target SoC device for highest security.

*A. Software Memory Complexity*

Assume that the GENIE would be allowed to insert a bitstream manipulator tool that can manipulate the configuration bitstream. The bitstream to be manipulated contains an application core design together with the *SUC* templates. The software GENIE contains:

- A Configuration Bitstream Manipulation Tool (BMT)
- Mappings storage: such as the class of optimal S-Boxes

The memory complexity of the GENIE is described as follows:

$$SW_c = C_{BMT} + C_{SM}$$

Where $C_{BMT}$ denotes the memory complexity of the bitstream manipulator tool which need to store all templates address locations. $C_{BMT}$ is relatively small in most cases. $C_{SM}$ represents the major memory complexity of the stored mappings. Each S-Box consumes 64 bits, for example storing the total set of optimal involutive S-Boxes in the software GENIE requires $C_{SM} = 8.90625$ *Mbits*.

*B. Hardware Complexity*

When the GENIE is realized inside a system controller as that of SmartFusion®2 SoC FPGA, the overhead hardware complexity is expected to be negligible since all required actions can be realized by BMT.

*C. Time Complexity*

During the personalization, the GENIE selects random mappings and then manipulates the configuration bitstream accordingly. This requires getting random number from the TRNG. The required number of bits from the TRNG for *I-SUC* and *NI-SUC* are $TRNG_{I-SUC}$ and $TRNG_{NI-SUC}$ respectively:

$$\begin{cases} TRNG_{I-SUC} = 16 \times \log_2\left(2^{17.15}\right) + \log_2\left(|RKSA_I|\right) \\ TRNG_{NI-SUC} = 16 \times \log_2\left(2^{20.4}\right) + \log_2\left(|RKSA_{NI}|\right) \end{cases}$$

This results with:

$TRNG_{I-SUC} = 156$ Bytes and $TRNG_{NI-SUC} = 170$ Bytes

It is expected that, the TRNG can generate the required number of bytes in real-time. Otherwise, such small number of random bits can be generated and stored in a dead-time before running the personalization process to save latency time in the enrollment phase.

## XI. Conclusion

A novel concept allowing self-creation of unknown cipher modules in SoC NV-FPGAs is presented. Such created so called Secret Unknown Cipher (SUC) converts the hosting device physically into a non-replaceable or hard to clone unit. The concept is based on pre-compiled cipher templates in the FPGA's bitstream to avoid design-rules violations. Two cipher classes with cardinalities exceeding $2^{1000}$ are proposed. A non-involutive and an involutive cipher classes. A prototype implementation in Microsemi SmartFusion®2 SoC FPGA technology is evaluated for both ciphers. The proposed concept shows also the feasibility of self-creation process of unknown functions as a paradigm towards future trends in a technology-oriented cryptography. The attained sample practical security levels and area complexities are very promising. Even when the required FPGA infrastructure for such techniques are not available today, it is shown that the necessary FPGA-changes seem to be feasible in future NV-FPGA technologies. Many other new promising application proposals deploying the same new paradigm are in progress.